\documentclass[reprint,aps,prab,superscriptaddress,amsmath,amssymb]{revtex4-1}

\usepackage{graphicx}
\usepackage{siunitx} 
\usepackage[super]{nth} 
\usepackage{amsfonts}

\usepackage[capitalize]{cleveref} 

\def\D{\mathrm{d}}
\newcommand{\fs}{\femto\second}
\newcommand{\pC}{\pico\coulomb}

\newcommand{\rsq}{$1-R^2$}

\begin{document}


\title{Temporal X-ray Reconstruction using Temporal and Spectral Measurements at LCLS}


\author{Florian Christie}
  \email{florian.christie@desy.de}
\affiliation{%
 Deutsches Elektronen-Synchrotron, 22607 Hamburg, Germany
}
\affiliation{Universit\"at Hamburg, 22607 Hamburg, Germany.}

\author{Alberto Andrea Lutman}
  \affiliation{
   SLAC National Accelerator Laboratory, Menlo Park, California 94025, USA
  }%


\author{Yuantao Ding} \author{Zhirong Huang} 
\author{Vatsal A. Jhalani}
\thanks{Present address: California Institute of Technology, Pasadena, California 91125, USA}

 \author{Jacek Krzywinski} \author{Timothy John Maxwell} \author{Daniel Ratner}
\affiliation{
 SLAC National Accelerator Laboratory, Menlo Park, California 94025, USA
}%
\author{Juliane R\"onsch-Schulenburg}%
\author{Mathias Vogt}
\affiliation{%
 Deutsches Elektronen-Synchrotron, 22607 Hamburg, Germany
}%


\date{\today}

\begin{abstract}
Transverse deflecting structures (TDS) are widely used in accelerator physics to measure the longitudinal density of particle bunches. When used in combination with a dispersive section, the whole longitudinal phase space density can be imaged. At the Linac Coherent Light Source (LCLS), the installation of such a device downstream of the undulators enables the reconstruction of the X-ray temporal intensity profile by comparing longitudinal phase space distributions with lasing on and lasing off. However, the resolution of this TDS is limited to around \SI{1}{\femto\second} rms (root mean square), and therefore, it is not possible to resolve single self-amplified spontaneous emission (SASE) spikes within one X-ray photon pulse. 
By combining the power spectrum from a high resolution photon spectrometer and the temporal structure from the TDS, the overall resolution is enhanced, thus allowing the observation of temporal, single SASE spikes. The combined data from the spectrometer and the TDS is analyzed using an iterative algorithm to obtain the actual intensity profile. 
In this paper, we present some improvements to the reconstruction algorithm as well as real data taken at LCLS.
\end{abstract}


\maketitle

\section{Introduction}
Transverse deflecting structures (TDS) are used to time-resolve the electron bunch phase spaces downstream of an X-ray free-electron laser (FEL) undulator line to measure the electron bunch energy losses induced by the lasing process~\cite{Behrens2014}.
The electron bunch time-resolved losses match the emitted X-ray temporal profile~\cite{ding_xtcav}, and therefore TDSs are routinely used as diagnostics to measure the X-ray pulse profiles~\cite{Behrens2014}.
However, the limited resolution of a measurement using a TDS imposes an upper limit on the resolution of the temporal reconstruction of photon pulses. As the resolution of an X-band TDS used at an X-ray FEL is typically limited to around \SI{1}{\fs} rms (root mean square), single self-amplified spontaneous emission (SASE) spikes, typically in the range of \SIrange{0.1}{3}{\fs}, within one photon pulse can often not be resolved. However, the exact knowledge of the temporal structure of SASE radiation is interesting for applications such as ``ghost imaging''~\cite{PhysRevX.9.011045}.

By combining the power spectrum from a high resolution photon spectrometer~\cite{zhu_spectrometer} and the temporal structure from the TDS, the overall resolution can be enhanced, thus allowing the observation of temporal, single SASE spikes in the X-ray range. The combined data from the spectrometer and the TDS is analyzed using an iterative algorithm to obtain an estimate of the actual intensity profile. This iterative reconstruction algorithm is published in~\cite{Christie_Temporal_X-ray,Christie_PhD}.

In the following, we will discuss some improvements and adjustments to this iterative reconstruction algorithm that are necessary to analyze real data due to the spectrometer resolution of \SI{0.2}{\electronvolt} during the measurements. 

\section{Adjustments to iterative reconstruction algorithm}\label{sec:Iterative_Reconstruction_Algorithm}
The iterative reconstruction algorithm is described in detail in~\cite{Christie_Temporal_X-ray,Christie_PhD}. 
The blurred temporal profile $\tilde{P}(t)$ measured by a TDS and the blurred power spectrum $\tilde{\mathcal{P}}(\omega)$ measured by a spectrometer are the measured data used by the reconstruction algorithm to retrieve the actual pulse profile.
The mechanism by which the finite TDS resolution blurs the actual temporal intensity profile $P(t)$ is modeled by a convolution with a Gaussian $G(t)$ of fixed standard deviation $R_t$
\begin{equation}
\left(P\ast G\right)(t) = \tilde{P}(t).
\end{equation}
In contrary to~\cite{Christie_Temporal_X-ray}, in this paper we assume that also the power spectrum $\mathcal{P}(\omega)$ measured by the spectrometer has a limited resolution $R_\omega$. This process is also modeled by a convolution with a Gaussian $G(\omega)$ of width $R_\omega$
\begin{equation}
\left(\mathcal{P}\ast G\right)(\omega) = \tilde{\mathcal{P}}(\omega).
\end{equation}
The blurred temporal profile $\tilde{P}(t)$ and the blurred power spectrum $\tilde{\mathcal{P}}(\omega)$ are the starting points for the algorithm.

\subsection{Linearly chirped base functions}
The electric field of the photon pulse to be approximated is modeled as a sum of in principle arbitrary base functions $B_j(t)$ in time with varying complex coefficients $a_{j,m}$, where 
$m$
is the iteration step,
\begin{equation} \label{eq:timebase_general}
F_m(t) = \sum_{j=1}^{n} a_{j,m} B_j(t) 
\end{equation}
as is the field in the frequency domain
\begin{equation} \label{eq:freqbase_general}
\mathcal{F}_m(\omega) = \sum_{j=1}^{n} a_{j,m}\mathcal{B}_j(\omega)
,
\end{equation}
where $\mathcal{B}_j(\omega)$ are the base functions in the frequency domain.

The Gaussian base functions described in~\cite{Christie_Temporal_X-ray} are not chirped in time as suggested by~\cite{Bonifaccio_TimeDomain,kim_huang_lindberg_2017,Kim_Temporal}. To accommodate this we introduce an arbitrary linear chirp factor $\beta_j$ to the base functions
\begin{equation}\label{eq:Linearly_Chirped_Time}
B_j(t) 
= 
\left(\frac{1}{\sqrt{2 \pi}\sigma_j}\right)^\frac{1}{2} 
e^{- \frac{\left( t-t_j\right) ^2}{4 \sigma_j^2} }
e^{i \omega_j t}
e^{i \left( t-t_j\right) ^2 \beta_j},
\end{equation}
where $\sigma_j$ is the width of the Gaussians centered at times $t_j$, and the $\omega_j$ can be initially calculated based on the energy profile of the electron phase space. For a linearly chirped electron bunch, we can for example set $\omega_j = \omega_0 + 2  \frac{\gamma_j- \gamma_0}{\gamma_0} \omega_0$~\cite{Krinsky_freq_chirped}, where $\omega_0$ is the main radiation frequency created by electrons with an energy of $\gamma_0$ and $\gamma_j$ is the mean energy of the electrons around $t_j$.
Otherwise, they are initialized as $\omega_j=\omega_0$. 

These base functions are chosen in a way that 
\begin{equation}\label{eq:integral_linearly_chirped}
\begin{split}
\int_{-\infty}^{\infty} \left| B_j(t) \right|^2 \,\D t& =
\int_{-\infty}^{\infty} \frac{1}{\sqrt{2 \pi} \sigma_j}\cdot 
e^{-\frac{\left(t-t_j\right)^2}{2 \sigma_j^2}} \,\D t\\
& =1.
\end{split}
\end{equation}

By setting $C_j=\frac{1}{4 \sigma_j^2}-i \beta_j$, \cref{eq:Linearly_Chirped_Time} becomes
\begin{equation}\label{eq:Linearly_Chirped_Time2}
B_j(t) 
= 
\left(\frac{1}{\sqrt{2 \pi}\sigma_j}\right)^\frac{1}{2} 
e^{- \left( t-t_j\right) ^2 C_j }
e^{i \omega_j t}
\end{equation}
and we obtain the base functions in frequency domain 
\begin{equation}\label{eq:Linearly_Chirped_Freq}
\mathcal{B}_j(\omega) 
= 
\left(\frac{1}{\sqrt{2 \pi}\sigma_j}\right)^\frac{1}{2}
\frac{1}{\sqrt{2 C_j}}
e^{- \frac{ \left(\omega-\omega_j \right) \left( -4 i C_j t_j + \omega+\omega_j \right) }{4 C_j} }.
\end{equation}

Following~\cite{kim_huang_lindberg_2017,Kim_Temporal} we set $\beta_j=-\frac{1}{4 \sqrt{3} \sigma_j^2}$ for all base functions.

The iteration process and the minimizing is described in detail in~\cite{Christie_Temporal_X-ray}. The only change applied to the algorithm is that instead of the actual power spectrum $\mathcal{P}(\omega)$ the blurred power spectrum $\tilde{\mathcal{P}}(\omega)$ is used for the projected spectrum 
for the real data taken at the Linac Coherent Light Source (LCLS).

\subsection{Testing of the algorithm using a blurred power spectrum}\label{sec:Testing}
Similar to~\cite{Christie_Temporal_X-ray} the algorithm was tested using simulation data of LCLS at \SI{1.5}{\nano\meter} with a blurred power spectrum using a resolution of $R_\omega=\SI{0.2}{\electronvolt}$ rms. 
This value was chosen to match the resolution of the spectrometer used for measurements at LCLS. 
Therefore, the same tests as in~\cite{Christie_Temporal_X-ray} are conducted and showcased in the following using a blurred power spectrum, $\left(\mathcal{P}\ast G\right)(\omega) = \tilde{\mathcal{P}}(\omega)$ with $R_\omega=\SI{0.2}{\electronvolt}$.

Simulations of the SASE process in the LCLS undulators at \SI{1.5}{\nano\meter} are conducted using the 1D leap-frog algorithm developed by Z. Huang~\cite{kim_huang_lindberg_2017}. 
Single FEL simulations were run for bunch charges of \SI{20}{\pico\coulomb}, \SI{40}{\pico\coulomb}, and \SI{150}{\pico\coulomb} charge, each resulting in a different X-ray pulse length. 
Further testing on the code is done using the intensity profiles of 50 different SASE shots right before saturation (using different initial seeds due to the statistical nature of SASE) for a charge of \SI{40}{\pico\coulomb}.

\begin{figure}[tbh]
   \centering
   \includegraphics[width=0.95\columnwidth]{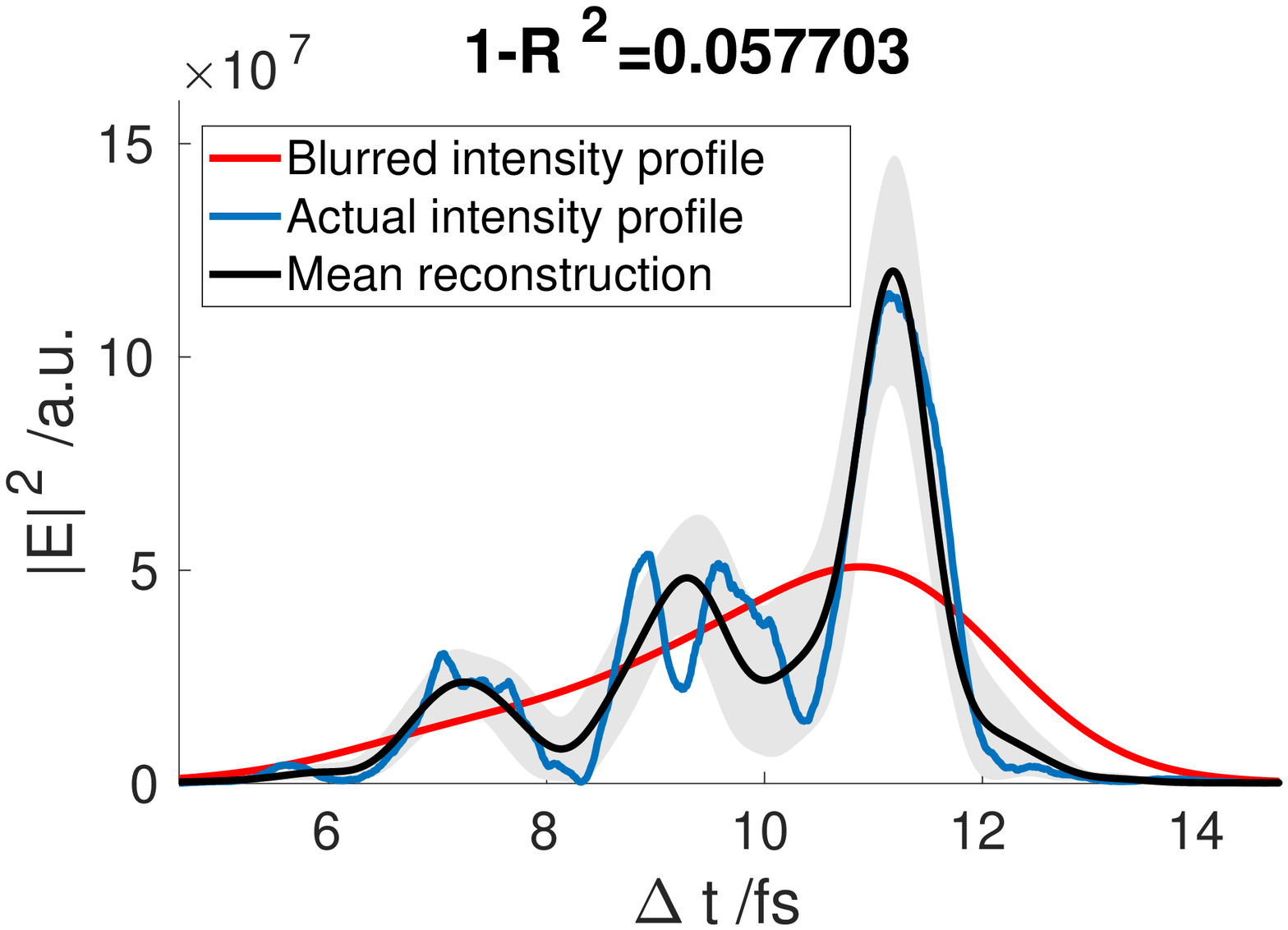}
   \caption{Reconstruction for \SI{20}{\pico\coulomb}, $R_t=\SI{1}{\femto\second}$.}
   \label{fig:20pCresults}
\end{figure}

\begin{figure}[tbh]
   \centering
   \includegraphics[width=0.95\columnwidth]{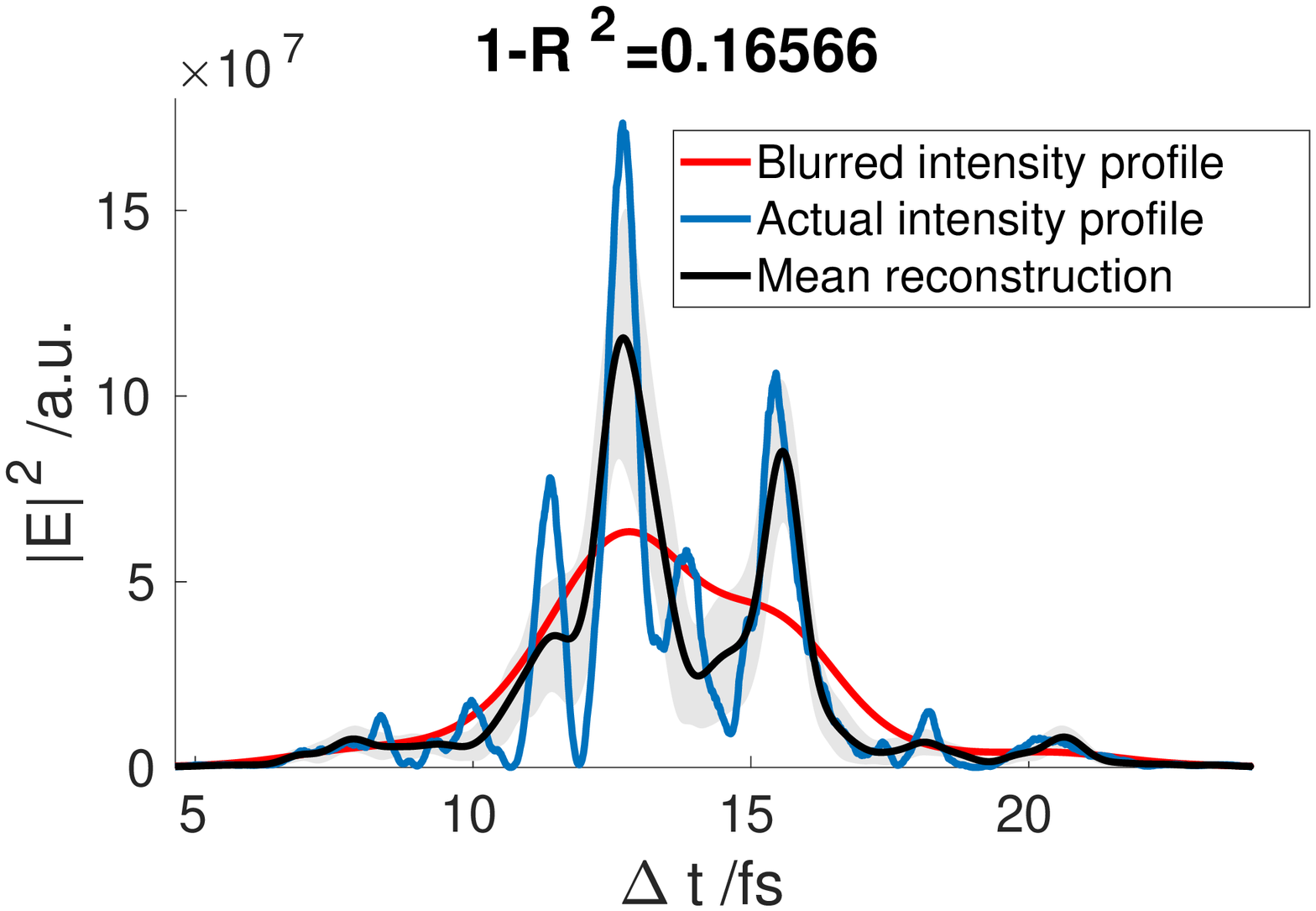}
   \caption{Reconstruction for \SI{40}{\pico\coulomb}, $R_t=\SI{1}{\femto\second}$.}
   \label{fig:40pCresults}
\end{figure}

\begin{figure}[t]
   \centering
   \includegraphics[width=0.95\columnwidth]{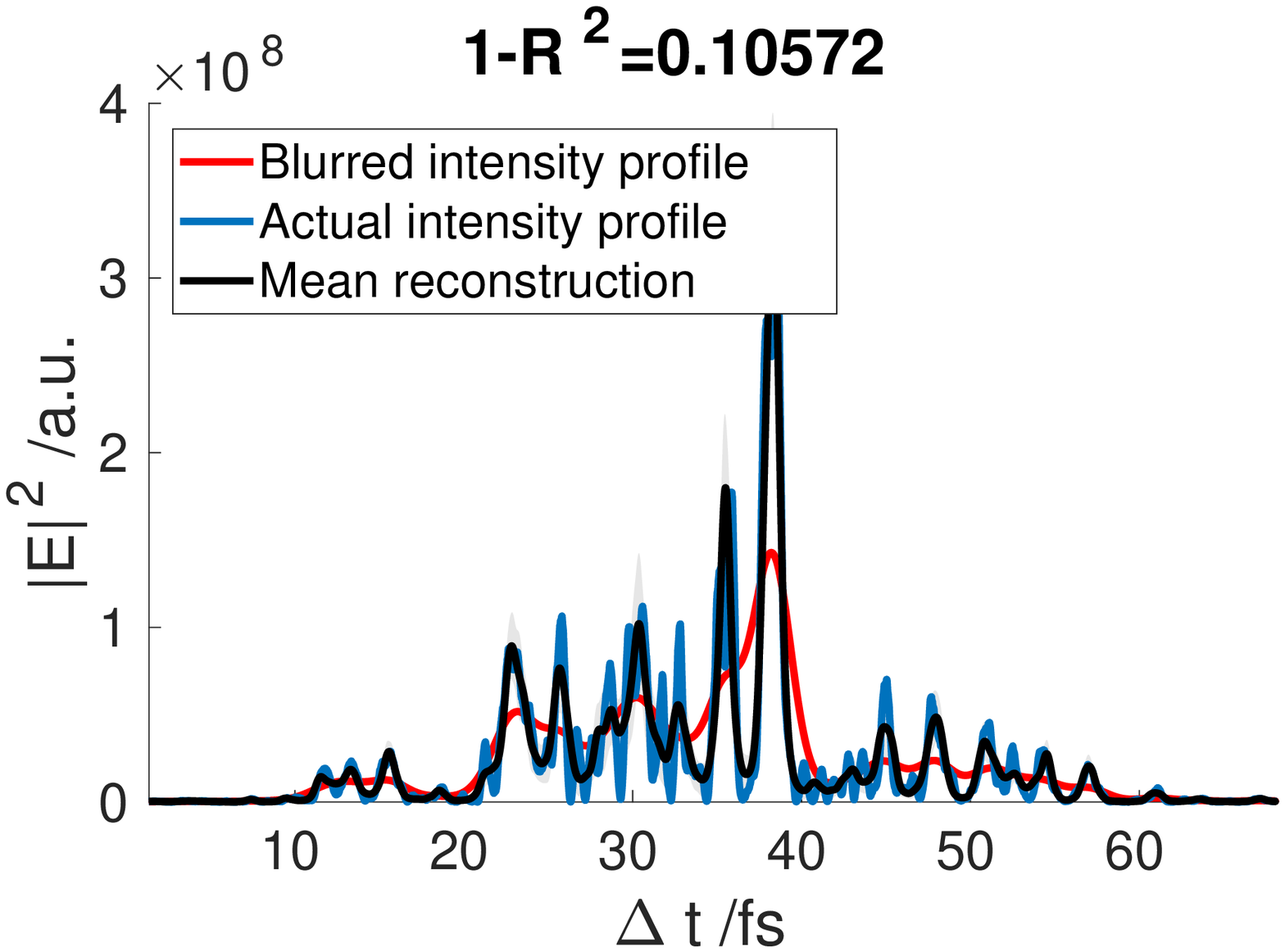}
   \caption{Reconstruction for \SI{150}{\pico\coulomb}, $R_t=\SI{1}{\femto\second}$.}
   \label{fig:150pCresults}
\end{figure}

For each shot 50 reconstructions using different initial base functions are carried out. To obtain the final reconstruction, these 50 reconstructions are then averaged. Figures~\ref{fig:20pCresults} to~\ref{fig:150pCresults} show the results for the different bunch charges: the blue solid line is the actual intensity profile from the simulation $P(t)$, the red solid line is the blurred intensity profile $\tilde{P}(t)$ using a resolution of $R_t=\SI{1}{\femto\second}$ in these cases. 
The black solid line is the mean of the 50 reconstructions surrounded by a light gray shaded area that is one standard deviation.
As can be observed, the reconstruction algorithm is still able to retrieve the actual intensity profile very similar to the cases without spectral blurring shown in~\cite{Christie_Temporal_X-ray}. The \rsq\ values are all in the same region as in~\cite{Christie_Temporal_X-ray} and also the capabilities and limitations of the algorithm remain the same as before. 

\begin{figure}[tbh]
   \centering
   \includegraphics[width=0.95\columnwidth]{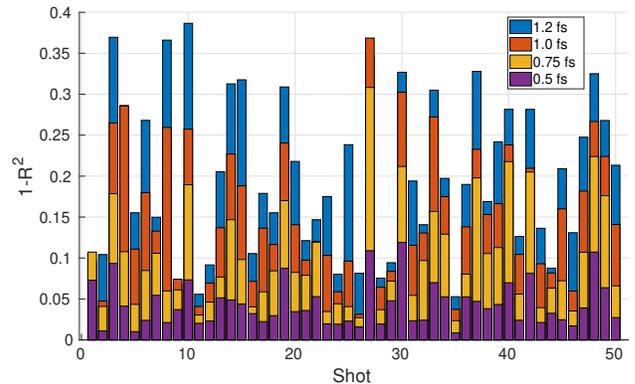}
   \caption[Reconstruction of 50 different SASE shots for \SI{40}{\pC} bunches with a spectral resolution of \SI{0.2}{\electronvolt}]{Results of the reconstruction of 50 different SASE shots for \SI{40}{\pC} bunches for different TDS resolutions shown in the legend. The $1-R^2$ value is plotted against the shot number. The power spectrum is blurred using a Gaussian blurring of $R_\omega=\SI{0.2}{\electronvolt}$.}
   \label{fig:40pC_50shots_blurred_spectrum}
\end{figure}

\Cref{fig:40pC_50shots_blurred_spectrum} shows the \rsq\ values for the 50 different shots with a charge of \SI{40}{\pC}. Comparing to~\cite{Christie_Temporal_X-ray} it can be seen, that for every shot the \rsq\ value is in the same order for the perfect and the blurred spectrum, respectively. It can therefore be concluded, that the algorithm performs equally well for both spectral measurements.

In summary, the testing results show that the algorithm excels at reconstructing single, isolated spikes and struggles to resolve multiple, dense spikes individually. The existence of adjoining spikes of similar height is retrieved, although the peak power cannot always be correctly determined. 

\section{Iterative reconstruction algorithm applied to measurement data taken at LCLS}
A dedicated machine development shift recorded data to be analyzed using the iterative reconstruction.

The measurements were taken at a beam energy of \SI{4}{\giga\electronvolt} resulting in a wavelength of $\sim\SI{1.7}{\nano\meter}$ or a photon energy of $\sim\SI{730}{\electronvolt}$.
The initial charge at the cathode was \SI{40}{\pC}, which is later collimated to either 20pC or 30pC prior to the undulator. Using energy collimators in the first bunch compressors, electrons with high energy deviations from the reference energy were truncated~\cite{PhysRevAccelBeams.19.100703}. An overview of the parameters for the measurements shown in the following can be found in \cref{tab:par_overview_LCLS_meas}.

\begin{table}[!htb]
\centering
\caption{Parameter Overview of Measurements for Iterative Reconstruction Algorithm at LCLS}
\label{tab:par_overview_LCLS_meas}
\begin{tabular}{|c|cc|}
\hline
Charge at undulator & \SI{30}{\pC}            & \SI{20}{\pC}            \\ \hline
Bunch length $\sigma_t$          & \SI{5}{\fs}            & \SI{3}{\fs}             \\
Peak current $I_{e}$          & \SI{2.2}{\kilo\ampere}  & \SI{2.5}{\kilo\ampere}  \\
TDS deflecting voltage $V$           & \SI{80}{\mega\volt}     & \SI{80}{\mega\volt}     \\
Temporal resolution $R_t$               & \SI{1.2}{\fs}           & \SI{1.0}{\fs}           \\
Spectral resolution $R_{\omega}$        & \SI{0.2}{\electronvolt} & \SI{0.2}{\electronvolt} \\ \hline
\end{tabular}
\end{table}

\begin{figure}
   \includegraphics[width=0.95\columnwidth]{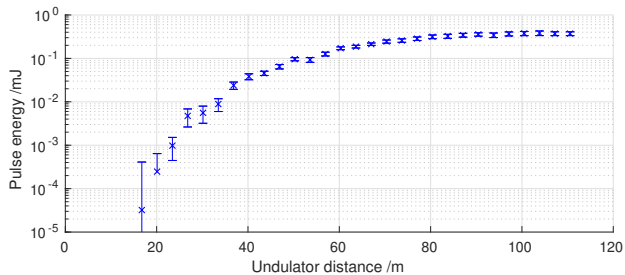}
   \caption{Gain curve measured at LCLS in the course of recording data for the reconstruction algorithm showing the gas detector signal over the distance traveled in the undulators. The error bars indicate one standard deviation. The beam was kicked behind the \nth{20} undulator (at $z=\SI{67}{\meter}$) to sustain the unsaturated condition of the bunches, but still provided sufficient signal for the spectrometer.\label{fig:Gain_Curve_LCLS_170731}}
\end{figure}

As the reconstruction has to be performed using bunches that are not yet saturated or close to the saturation point~\cite{maxwell_reconstruction}, a gain curve was recorded, see \cref{fig:Gain_Curve_LCLS_170731}. For this reason the beam was kicked behind the \nth{20} undulator (at $z=\SI{67}{\meter}$) to suppress lasing in the downstream undulators. 
The orbit kick both disrupts overlap between the x-ray waves and electron bunch and degrades electron bunching, interrupting the lasing process~\cite{gain_length_lcls}.
The chosen configuration provided sufficient signal for the spectrometer to work and fulfilled the requirement of being close to the saturation point so that a meaningful reconstruction of the photon pulse power from the energy loss and the energy spread increase of the electron bunch can be performed~\cite{maxwell_reconstruction}. The orbit downstream of the undulator section was restored stable by using a closed three-bump~\cite{gain_length_lcls,Wille,fmueller_master}.

To obtain the photon pulse power the longitudinal phase space density of an electron bunch producing light (lasing on) has to be compared to one where the lasing process is suppressed (lasing off).
The longitudinal phase space densities are then divided into slices along the time dimension to get the time-dependent beam parameters such as the mean energy $E_{\text{on},\text{off}}(t_i)$, the energy spread $\sigma_{E_{\text{on},\text{off}}}(t_i)$, and the current $I(t_i)$ in each time slice $t_i$~\cite{Behrens2014,maxwell_reconstruction}. The subscript denotes, whether the quantity was taken from a measurement with lasing on or lasing off. The influence of the FEL process on the bunch current is negligible, therefore, this quantity does not have a subscript.
For each time slice, the energy loss and the energy spread increase comparing the lasing-on and lasing-off measurement is then calculated
\begin{equation}
	\Delta E(t_i)=E_\text{on}(t_i)-E_\text{off}(t_i),
\end{equation}
\begin{equation}
\sigma_E(t_i)=\sqrt{\sigma_{E_\text{on}}^2(t_i)-\sigma_{E_\text{off}}^2(t_i)}.
\end{equation}
From these quantities the radiation power in each slice can be determined. When using the energy loss method, the radiation power in each slice is~\cite{ding_xtcav,Behrens2014,maxwell_reconstruction}
\begin{equation}\label{eq:energy_loss_method}
P(t_i)=\Delta E(t_i)\cdot\frac{I(t_i)}{e}.
\end{equation}
When using the energy spread method, the radiation power in each slice is~\cite{Behrens2014,maxwell_reconstruction,HUANG2008120}
\begin{equation}\label{eq:energy_spread_method}
P(t_i) \propto \sigma_E(t_i)^2\cdot I(t_i)^\frac{2}{3}.
\end{equation}
To determine the total radiation power an additional, independent measurement of the total pulse energy is necessary. This can for example be accomplished using a calibrated gas detector~\cite{gmd,gmd2,MOELLER2011S6}.

To see if the algorithm reconstructs the photon pulse correctly both, the power profile obtained from the energy loss and the energy spread method, are used as target power profiles for the reconstruction algorithm. By then comparing the reconstructed temporal power profiles one can observe the similarities and differences to check if even though the inputs might be slightly different the algorithm ends up with the same temporal power profile. 

In the following, the photon pulses obtained from the reconstruction using only the TDS and the corresponding reconstruction methods are displayed as dashed-dotted lines. The reconstructions using the energy loss method are plotted in black and those using the energy spread method are plotted in blue. 
50 different initial guesses serve as starting points for the reconstruction algorithm. 
These 50 reconstructions are averaged to obtain the final reconstructed photon pulses (solid lines). 
The mean reconstruction is surrounded by a dark gray and a light gray shaded area which is one standard deviation for the energy loss and the energy spread method, respectively.
For the reconstruction the width of the Gaussian base functions $\sigma_j$ was chosen to be \SI{0.1}{\fs} as well as their spacing 
$\Delta t=t_{j+1}-t_{j}$.

\subsection{Reconstruction of photon pulses from bunches with \SI{30}{\pC} charge}
We first apply the reconstruction algorithm the experimental data with a charge of \SI{30}{\pC} at the undulator entrance. 
The parameter overview can be found in the left column of \cref{tab:par_overview_LCLS_meas}.

\begin{figure}[tbh]
   \includegraphics[width=\columnwidth]{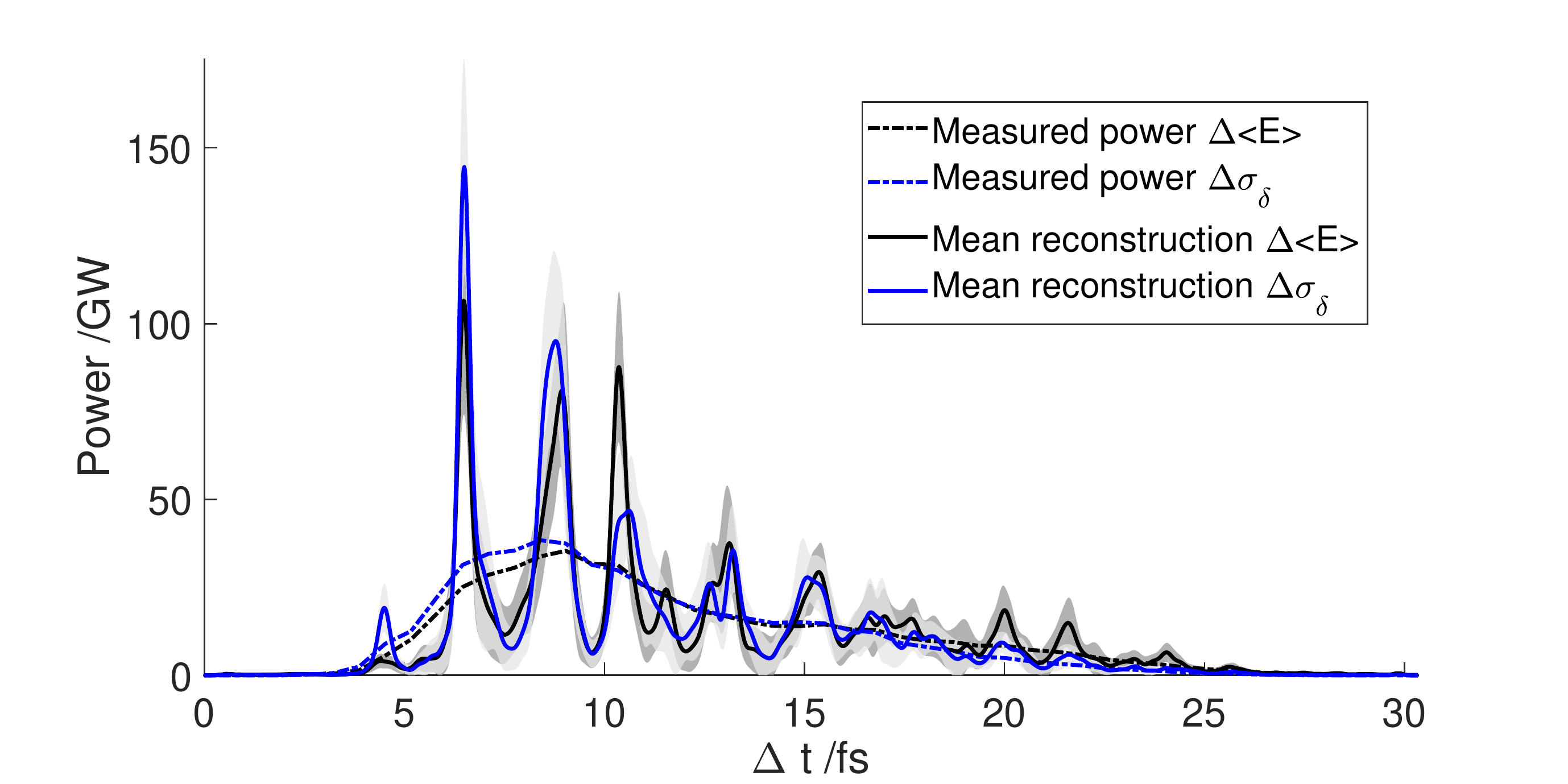}
   \caption[Photon pulse reconstruction at LCLS for \SI{30}{\pC}, Shot 1]{Reconstruction of a photon pulse measured at LCLS obtained using the iterative reconstruction algorithm, Shot 1. The total bunch charge at the undulator is \SI{30}{\pC}, the TDS resolution is \SI{1.2}{\fs}.   \label{fig:LCLS_run_1527_bsl_1523_shot_11}}
\end{figure}

\begin{figure}[tbh]
   \includegraphics[width=\columnwidth]{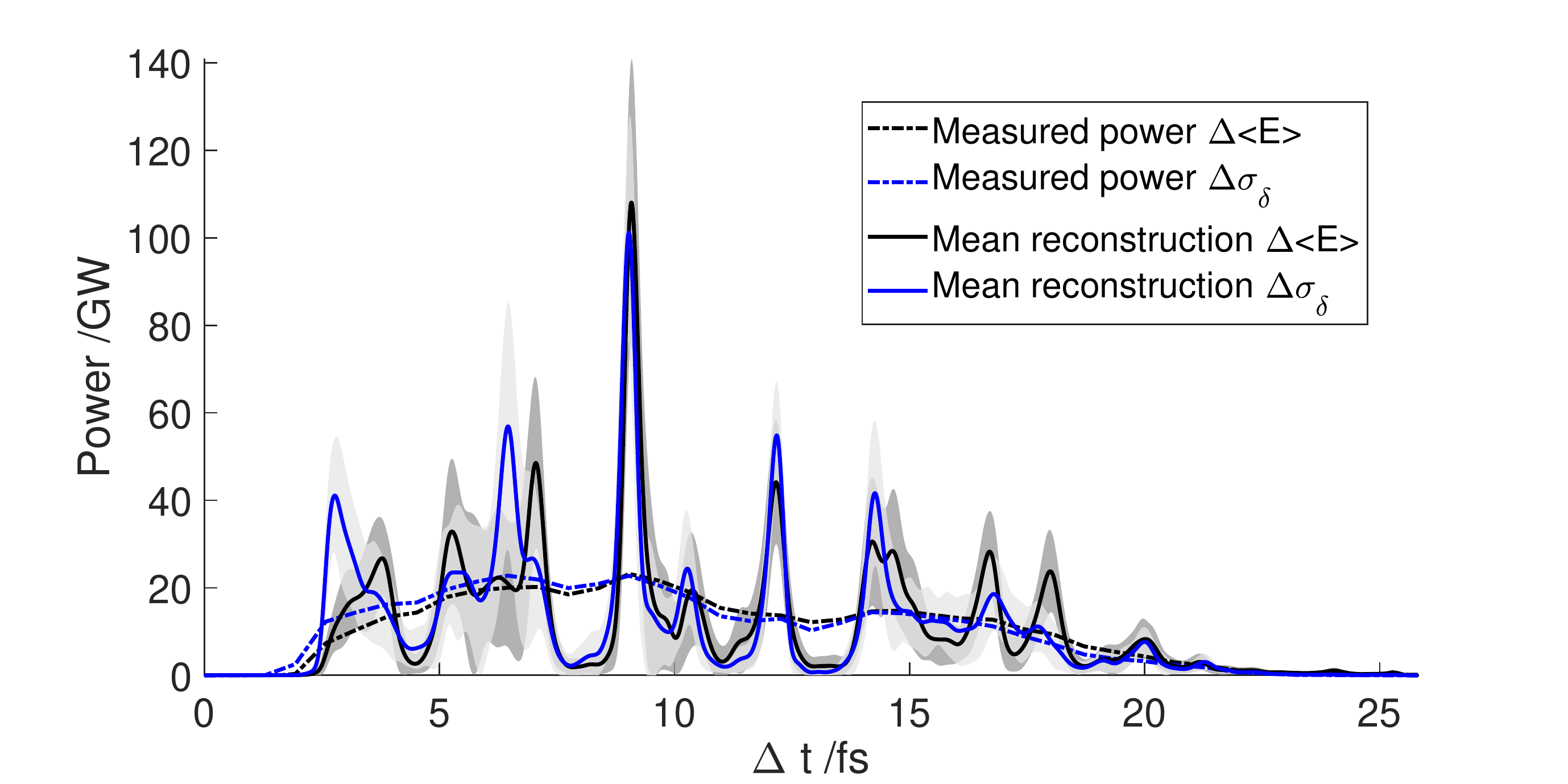}
   \caption[Photon pulse reconstruction at LCLS for \SI{30}{\pC}, Shot 2]{Reconstruction of a photon pulse measured at LCLS obtained using the iterative reconstruction algorithm, Shot 2. The total bunch charge at the undulator is \SI{30}{\pC}, the TDS resolution is \SI{1.2}{\fs}.   \label{fig:LCLS_run_1527_bsl_1523_shot_08}}
\end{figure}

\begin{figure}[t]
   \includegraphics[width=\columnwidth]{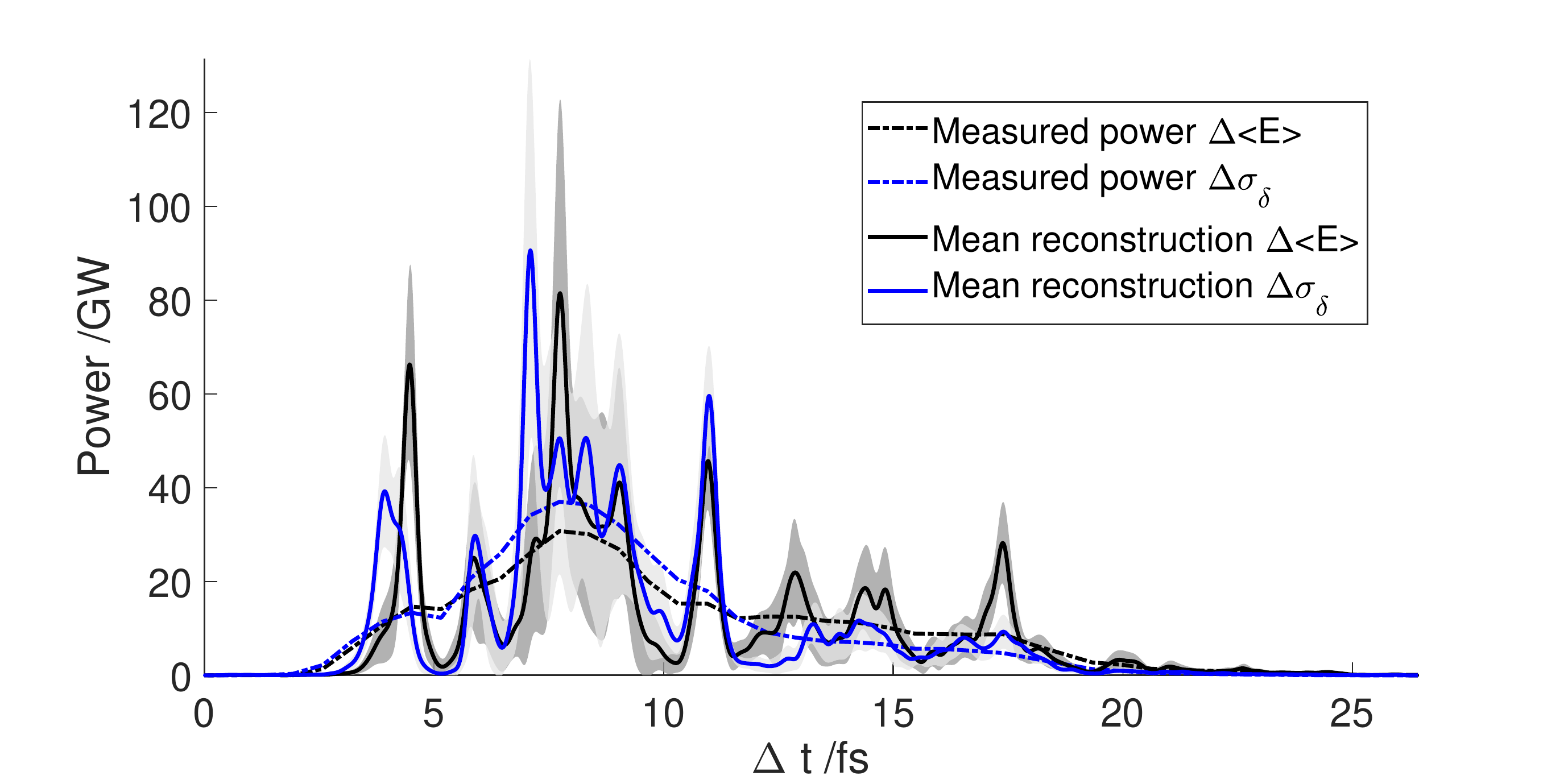}
   \caption[Photon pulse reconstruction at LCLS for \SI{30}{\pC}, Shot 3]{Reconstruction of a photon pulse measured at LCLS obtained using the iterative reconstruction algorithm, Shot 3. The total bunch charge at the undulator is \SI{30}{\pC}, the TDS resolution is \SI{1.2}{\fs}.   \label{fig:LCLS_run_1527_bsl_1523_shot_06}}
\end{figure}

Examples of reconstructed photon pulses can be found in 
\crefrange{fig:LCLS_run_1527_bsl_1523_shot_11}{fig:LCLS_run_1527_bsl_1523_shot_06}, demonstrating the capabilities and the limitations of the algorithm when applied to measured data.
As can be seen in these figures, the photon pulse obtained from the energy loss and the energy spread method differ slightly. 

For the power profiles in \cref{fig:LCLS_run_1527_bsl_1523_shot_11} the position of most of the larger SASE spikes is the same for both reconstructions. 
The maximum power of the highest spike is $\sim 1.36$ times higher if the energy spread method is used as the target profile. This is because the power obtained from the TDS reconstruction at the position of the spike is $\sim 1.30$ times higher if the energy spread method is used. 
The smaller SASE spikes agree very well in position and also in height. 
The reconstruction works well, as the main SASE spikes are separated, which according to Section~\ref{sec:Testing} facilitates the reconstruction for the algorithm.

The position and height of most of the main SASE spikes is also very similar for both target profiles shown in \cref{fig:LCLS_run_1527_bsl_1523_shot_08}. 
The maximum power of the main spike differs by less than \SI{7}{\percent}. 
There is a slight difference in the reconstruction on the left side of the main spike. 
Here, multiple spikes are very close to one another and the reconstruction algorithm reaches its limitations providing slightly different results for the two target profiles. 
It can be observed that the power profile contains two spikes between \SIrange[range-phrase ={ and }]{2}{4.5}{\fs} and three between \SIrange[range-phrase ={ and }]{4.5}{8}{\fs}. The position can be reconstructed, yet the exact maximum power for each single spike remains unknown. 
The isolated spikes in the region of \SIrange[range-phrase ={ and }]{10}{20}{\fs} can be retrieved efficiently by the algorithm.

\Cref{fig:LCLS_run_1527_bsl_1523_shot_06} shows a reconstruction where the algorithm reaches the limitations for reconstructing the main features of the photon pulse. These limitations were noted in Section~\ref{sec:Testing}.
In the central part of the photon pulse (between \SIrange[range-phrase ={ and }]{6}{10}{\fs}) the SASE spikes are too close to one another to be retrieved by the algorithm. 
Nonetheless, the smaller, isolated side peaks to the left and right of the central part are retrieved adequately by the algorithm.

\subsection{Reconstruction of photon pulses from bunches with \SI{20}{\pC} charge}
Secondly, bunches with a charge of \SI{20}{\pC} at the undulator entrance are used for the reconstruction. The settings are the same as in the previous section, but the bunches are truncated even further using energy collimators in the first bunch compressor. 
The resulting parameters can be found in the right column of \cref{tab:par_overview_LCLS_meas}.

\begin{figure}[tbh]
   \includegraphics[width=\columnwidth]{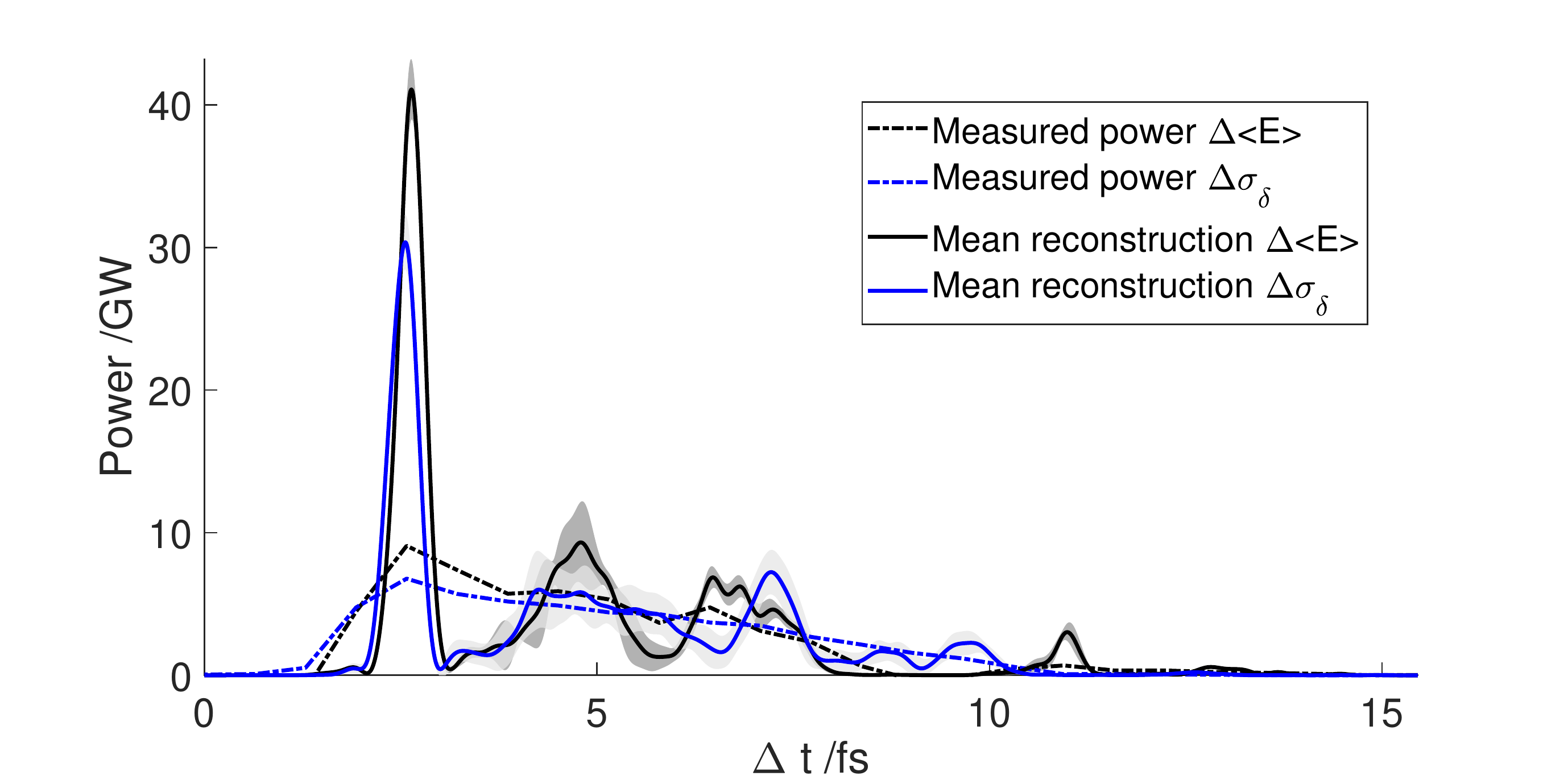}
   \caption[Photon pulse reconstruction at LCLS for \SI{20}{\pC}, Shot 1]{Reconstruction of a photon pulse measured at LCLS obtained using the iterative reconstruction algorithm, Shot 1. The total bunch charge at the undulator is \SI{20}{\pC}, the TDS resolution is \SI{1.0}{\fs}.   \label{fig:LCLS_run_1534_bsl_1536_shot_05}}
\end{figure}

\begin{figure}[tbh]
   \includegraphics[width=\columnwidth]{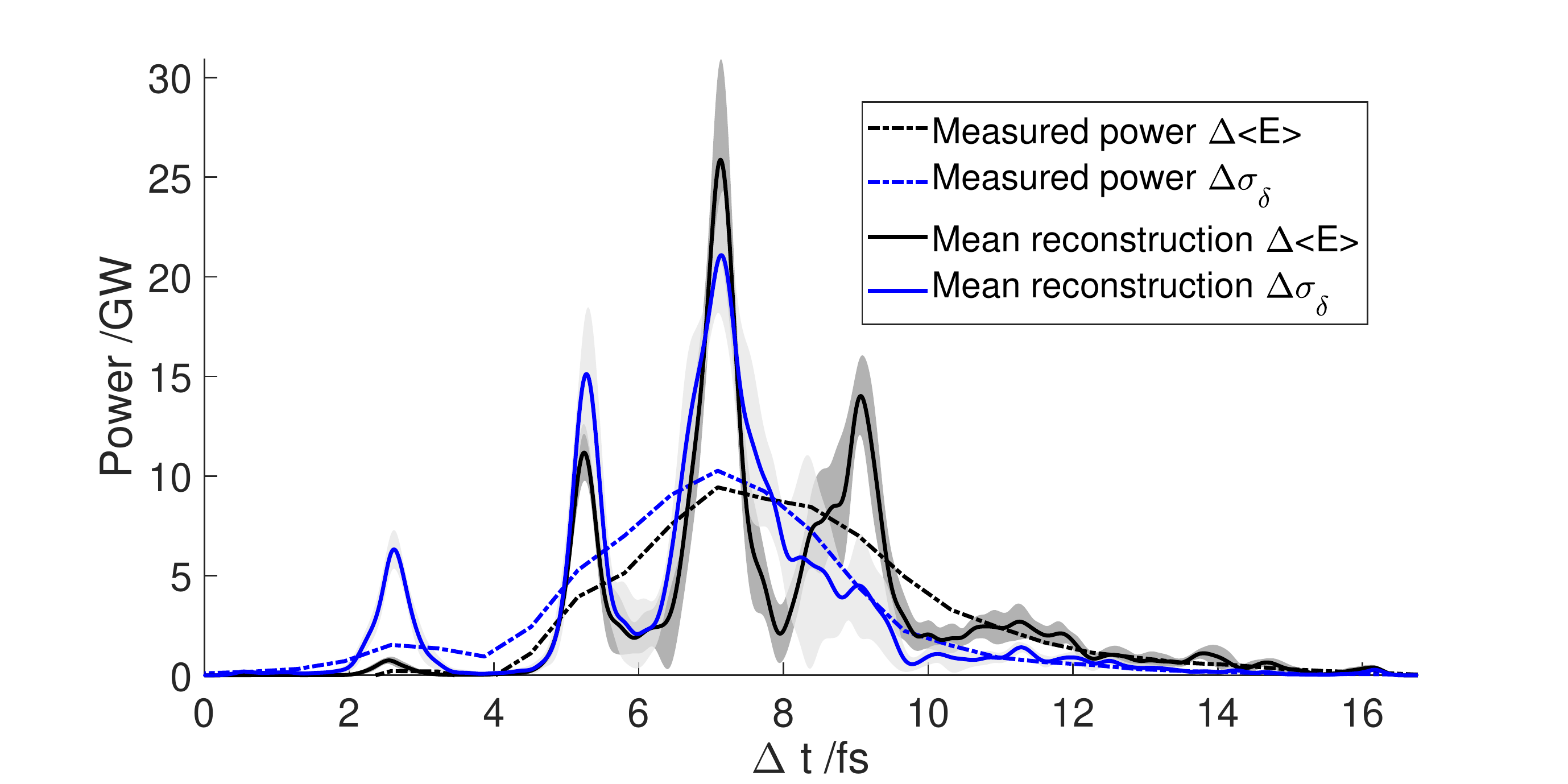}
   \caption[Photon pulse reconstruction at LCLS for \SI{20}{\pC}, Shot 2]{Reconstruction of a photon pulse measured at LCLS obtained using the iterative reconstruction algorithm, Shot 2. The total bunch charge at the undulator is \SI{20}{\pC}, the TDS resolution is \SI{1.0}{\fs}.   \label{fig:LCLS_run_1534_bsl_1536_shot_04}}
   \vspace{-4mm}
\end{figure}

\begin{figure}[!tbh]
   \includegraphics[width=\columnwidth]{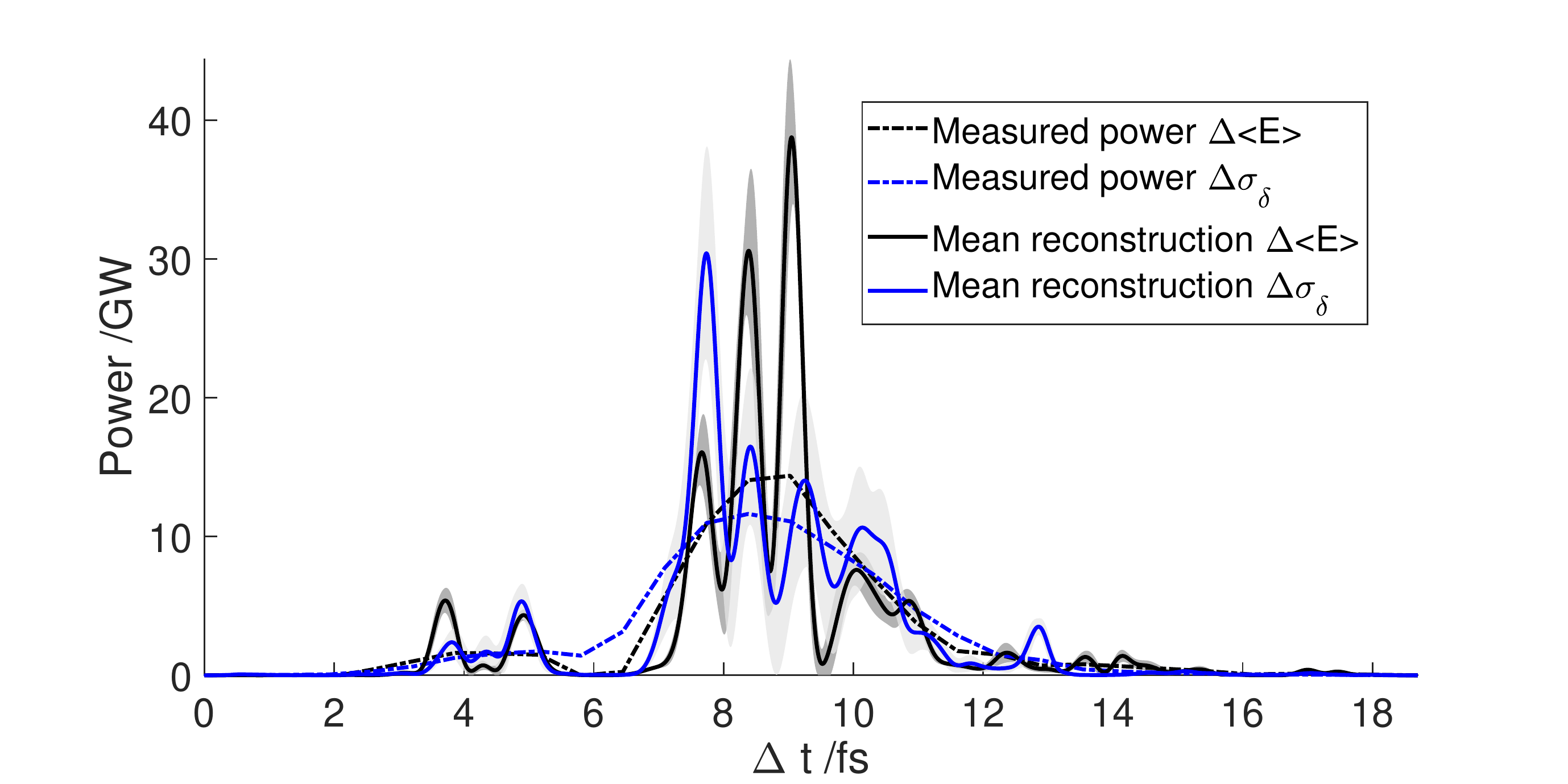}
   \caption[Photon pulse reconstruction at LCLS for \SI{20}{\pC}, Shot 3]{Reconstruction of a photon pulse measured at LCLS obtained using the iterative reconstruction algorithm, Shot 3. The total bunch charge at the undulator is \SI{20}{\pC}, the TDS resolution is \SI{1.0}{\fs}.   \label{fig:LCLS_run_1534_bsl_1536_shot_11}}
   \vspace{-4mm}
\end{figure}

\Crefrange{fig:LCLS_run_1534_bsl_1536_shot_05}{fig:LCLS_run_1534_bsl_1536_shot_11} show examples of reconstructed photon pulses for these settings demonstrating the capabilities and the limitations of the algorithm when applied to measured data. 

The reconstruction found in \cref{fig:LCLS_run_1534_bsl_1536_shot_05} shows a dominant spike at the beginning of the photon pulse. 
Both methods reconstruct the spike at \SI{2}{\fs} but with a different maximum power. 
With the energy difference method the power is $\sim 1.35$ times higher, in good agreement with the blurred power, which is $\sim 1.34$ times higher. 
The rest of the photon pulse consists of smaller, dense spikes which can only partly be retrieved by the algorithm. In the region between \SIrange[range-phrase ={ and }]{8}{10}{\fs} the photon pulse power reconstructed using the energy difference method is close to zero.
As a result the algorithm does not reconstruct any power in that region. 
On the contrary, the energy spread method yields power in this region which results in the reconstruction algorithm showing spikes here.
Hence, the difference in the reconstruction is not caused by instability in the algorithm, but rather by the difference in the two TDS analysis methods.

The second example for this setting can be found in \cref{fig:LCLS_run_1534_bsl_1536_shot_04}. The main spikes between \SIrange[range-phrase ={ and }]{6}{8}{\fs} and \SIrange[range-phrase ={ and }]{4}{6}{\fs} of the photon pulse are retrieved using both methods as starting points.
The small spike between \SIrange[range-phrase ={ and }]{2}{3}{\fs} of the photon pulse is higher if the energy spread method is used, as the photon pulse power reconstructed using this method is also higher. 
Several adjacent spikes in the region of \SIrange[range-phrase ={ to }]{8}{10}{\fs} cannot be clearly distinguished by the reconstruction algorithm.
Additionally, the power obtained by the energy difference method is higher in this region, leading to a higher power retrieved by the algorithm.

An example where the iterative algorithm did not reconstruct successfully can be found in \cref{fig:LCLS_run_1534_bsl_1536_shot_11}. Here, the central part of the photon pulse comprises many adjacent SASE spikes that cannot be distinguished by the algorithm. 
This is expected since the initial reconstruction using the two methods differs in the region of \SIrange[range-phrase ={ to }]{7}{11}{\fs} both in height and shape and thus, the algorithm ends up with different solutions in this region.


\section{Conclusions}
In this paper, the reconstruction algorithm published in~\cite{Christie_Temporal_X-ray} was improved and applied to real measurement data taken at LCLS. 

The results show how the reconstruction algorithm using both TDS and spectral information can improve the X-ray pulse profile reconstruction over the established method using only the TDS measurement. 
The reconstruction accuracy is currently limited by the achievable resolution of the TDS implemented at LCLS. For future upgrades an even better accuracy can be expected.
The algorithm excels at reconstructing single, isolated spikes. Multiple, adjacent spikes of similar height are more difficult to be retrieved individually.

\section*{Acknowledgments}
This work was supported by the U.S. Department of Energy (DOE) under Contract No. DE-AC02-76SF00515.

This is a pre-print of an article published in Scientific Reports. The final authenticated version is available online
at: \url{https://doi.org/10.1038/s41598-020-66220-5}.

\end{document}